\documentclass[apj]{emulateapj}

\usepackage{amsmath}
\usepackage{amsfonts}
\usepackage{amssymb}
\usepackage{color}

\usepackage{animate}

\shorttitle{New Limits on Polarized Power Spectra at 126 and 164 MHz}
\shortauthors{Moore, et al.}

\usepackage{natbib}
\newcommand{\Vector}[1]{\mathbf{#1}}
\newcommand{\D}[1]{\mathrm{d}#1}
\newcommand{\RNum}[1]{\uppercase\expandafter{\romannumeral #1\relax}}

\def\penn{1}
\def\ionFR{{\tt ionFR}}

\begin{document}

\title{Limits on Polarized Leakage for the PAPER Epoch of Reionization Measurements at 126 and 164 MHz}

\author{
  David F. Moore\altaffilmark{\penn},
  James E. Aguirre\altaffilmark{1},
  Saul A. Kohn\altaffilmark{1},
  Aaron R. Parsons\altaffilmark{2,8},
  Zaki S. Ali\altaffilmark{2},
  Richard F. Bradley\altaffilmark{3,4,5},
  Chris L. Carilli\altaffilmark{6,7},
  David R. DeBoer\altaffilmark{8},
  Matthew R. Dexter\altaffilmark{8},
  Nicole E. Gugliucci\altaffilmark{13},
  Daniel C. Jacobs\altaffilmark{9},
  Pat Klima\altaffilmark{4},
  Adrian Liu\altaffilmark{2,10},
  David H.~E. MacMahon\altaffilmark{8},
  Jason R. Manley\altaffilmark{11},
  Jonathan C. Pober\altaffilmark{12},
  Irina I. Stefan\altaffilmark{7},
  and
  William P. Walbrugh\altaffilmark{11}
}
\altaffiltext{1}{Dept. of Physics and Astronomy,
                  U. Pennsylvania,
                  Philadelphia, PA}
\altaffiltext{2}{Astronomy Dept.,
                  U. California,
                  Berkeley, CA}
\altaffiltext{3}{Dept. of Electrical and Computer Engineering,
                  U. Virginia,
                  Charlottesville, VA}
\altaffiltext{4}{National Radio Astronomy Obs.,
                  Charlottesville, VA}
\altaffiltext{5}{Dept. of Astronomy,
                  U. Virginia,
                  Charlottesville, VA}
\altaffiltext{6}{National Radio Astronomy Obs.,
                  Socorro, NM}
\altaffiltext{7}{Cavendish Lab.,
                  Cambridge, UK}
\altaffiltext{8}{Radio Astronomy Lab.,
                  U. California,
                  Berkeley, CA}
\altaffiltext{9}{School of Earth and Space Exploration,
                  Arizona State U.,
                  Tempe, AZ}
\altaffiltext{10}{Berkeley Center for Cosmological Physics,
                    U. California,
                    Berkeley, CA}
\altaffiltext{11}{Square Kilometer Array, South Africa Project,
                    Cape Town, South Africa}
\altaffiltext{12}{Dept. of Physics,
                    Brown University, 
                    Providence, RI}
\altaffiltext{13}{St. Anselm College,
                  Manchester, NH}
                    
\begin{abstract}

 Polarized foreground emission is a potential contaminant of attempts
 to measure the fluctuation power spectrum of highly redshifted 21 cm
 H{\sc i} emission from the Epoch of Reionization. Using the Donald
 C. Backer Precision Array for Probing the Epoch of Reionization, we present limits on the observed power spectra of all four
 Stokes parameters in two frequency bands, centered at 126 MHz
 ($z=10.3$) and 164 MHz ($z=7.66$), for a three-month observing
 campaign of a deployment involving 32 antennas, for which unpolarized power
 spectrum results have been reported at $z=7.7$ \citep{Parsons2014}
 and at $7.5 < z < 10.5$ \citep{Jacobs2014}.  The power spectra in this
 paper are processed in the same way as in those works, and show no
 definitive detection of polarized power. This nondetection is consistent 
 with what is known about polarized sources, combined with 
 the suppression of polarized power by
 fluctuations in the ionospheric rotation measure, which can strongly affect
 Stokes $Q$ and $U$. We are able to show that the net effect of polarized
 leakage is a negligible contribution at the levels of the limits
 reported by \citet{Parsons2014} and \citet{Jacobs2014}.

\end{abstract}

\maketitle

\section{Introduction}

Polarized emission at meter wavelengths is a potentially problematic
foreground for experiments seeking to measure the 21 cm power spectrum of the Epoch of
Reionization (EoR).  Smooth-spectrum sources occupy a
well-defined ``wedge'' in the cylindrical $(k_\parallel, k_\perp)$
space of the EoR power spectrum \citep[e.g.,][]{Morales2012,
  Pober2013}, but it has been understood for some time
\citep[e.g.,][]{Geil2011,Pen2009} that Faraday-rotated polarized
emission can contaminate the measurement of unpolarized EoR emission.
When mapped into the power spectrum, this Faraday-rotated polarized
emission generates power that mimics the high $k_{\parallel}$
emission of the EoR, scattering power into the otherwise clean EoR
window.  \citet[][hereafter M13]{Moore2013} simulated the effects of
the low-level forest of weak, polarized point sources leaking into the
full 3D EoR power spectrum.  M13 used the best measurements available
at the time, notably the existing (unpolarized) point source surveys
\citep{Hales1988,Cohen2004} and the few polarization measurements
below 200 MHz \citep{Pen2009,Bernardi2009}, and found that the level
of emission could be problematic for any experiment with $\sim1\%$
leakage from Stokes $Q \to I$.  More recent work, discussed further
below, suggests that the actual level of point source polarization is
lower than M13 assumed \citep{asad15}.  
It  may also be possible to address polarization leakage due to wide-field beams with modest calibration requirements, as demonstrated with the Murchison Widefield Array (MWA, a precursor of SKA, the Square Kilometer Array) \citep{sutinjo15mwa} and ongoing characterization of the SKA prototype antennas \citep{sutinjo15ska}.

Polarized emission may arise from diffuse (presumably Galactic)
emission and point sources.  Historically, the first measurements of
the polarization of diffuse emission in the Southern Hemisphere at
frequencies below 1.4 GHz were reported by \citet{mathewson64,mathewson65},
who used the Parkes telescope at 408 MHz and 0.8\arcdeg\ resolution to
follow on the previous work of
\citet{westerhout62}. \citet{spoelstra84} provided a summary of the
situation at frequencies between 408 and 1411 MHz, noting the low
rotation measures of this emission ($< 8~\rm{rad~m^{-2}}$) with a
polarization fraction of $\sim35\%$ near 1.4 GHz and notable
depolarization at lower frequencies.  The emission seems to be largely
due to nearby sources ($\sim$450 pc), and fluctuates on scales from 1
- 10\arcdeg.  More recent analysis of the angular power spectrum of
diffuse, polarized fluctuations by \citet{laporta06} suggests that the
fluctuations continue to smaller scales as a power law, depending on
frequency.  \citet[][hereafter B13]{Bernardi2013}, who conducted a
2400 square degree survey using the MWA at
189 MHz, and \citet[hereafter Je14]{Jelic2014}, both reported
significant amounts of weakly polarized emission on angular scales
between a few degrees for B13 and a few tens of arcminutes for
Je14. The rotation measure (RM) $|\Phi| \le 25~\rm{rad~m}^{-2}$ in all
cases.

Regarding point sources, the number of bright sources below 200 MHz
reported in the literature is small (although it should also be pointed out that only small parts of the southern and northern sky have been surveyed for such sources). Of particular interest in
determining the polarization contamination is the linearly polarized
fraction $p = \sqrt{(Q^2 + U^2)/I^2}$ of sources (where $I$, $Q$, and
$U$ are the Stokes parameters), and, in the absence of low-frequency
measurements, how this scales from higher frequencies.  B13 detected
one polarized point source at 189 MHz, PMN J0351-2744, with a
polarization fraction $p = 0.02$ and polarized flux of 320 mJy.  At
somewhat higher frequencies (350 MHz), \citet{Giessuebel2013} studied
sources behind M31 and found that $p_{\rm350~MHz} = 0.14~p_{\rm
  1.4~GHz}$, with typical values of a few percent at 350 MHz.  A study
of the depolarization of point sources by \citet{Farnes2014} showed a
systematic trend for depolarization of steep-spectrum point sources as
frequency decreased, resulting in very low polarization fractions ($p
\ll 0.01$) below 300 MHz.  In a measurement of M51 using LOFAR between
115 and 175 MHz, \citet{mulcahy14} detected 6 background sources with
Stokes $I$ between 44 and 1500 mJy.  They detect two ``sources'' with
relatively large polarized fractions (0.026 and 0.029), but these are
the lobes of radio galaxies with unpolarized cores.  If one were to
include the flux from the core, thus giving the total polarized
fraction integrated over the galaxy, this would be lower. For the
unresolved sources, the polarized fractions are all
$<2.8\times10^{-3}$.  Depolarization from 1.4 GHz ranged between 0.03
and 0.2 for the six sources.  Importantly, \citet{mulcahy14} had very
high angular resolution (20-25\arcsec) relative to B13 and the Donald C. Backer Precision Array for Probing the Epoch of Reionization (PAPER), thus
minimizing beam depolarization effects from integrating over sources.

It is worth noting that the reported rotation measures for point
sources vary over the range of values reported in the all-sky RM map
of \citet{Oppermann2012}, whereas the diffuse emission typically shows
low RM.  This is consistent with the interpretation of this emission
as nearby diffuse Galactic synchrotron.  As discussed in
\citet{Jelic2010}, polarized synchrotron generated from within a
magnetized, ionized plasma will be depolarized, and will further show
structure in the RM spectrum of the source.  The result
is a lower apparent Faraday depth.  By contrast, extragalactic point
sources see a large Faraday column through the Galaxy, resulting in a
high Faraday depth.

The goal of this paper is to determine, from measured polarization power spectra, the extent of possible polarized leakage on the upper limits on the EoR power spectrum previously reported from PAPER
in \citet[][hereafter P14]{Parsons2014} and
\citet[][hereafter J15]{Jacobs2014}. We begin in
Section \ref{sec:PolarizationReview} by reviewing key points of the
effect of polarization on the EoR power spectrum. We describe the
dataset and processing steps in Section \ref{sec:DataProcessing}, and
present power spectra from two frequency bands in all four Stokes
parameters in Section \ref{sec:Results} along with implications of
these measurements for the level of polarized leakage into the 21 cm
EoR power spectrum.  We conclude in Section \ref{sec:Conclusion}.

\section{The Problem of Polarization}
\label{sec:PolarizationReview}

In this section, we will briefly review the problem of polarization
for EoR power spectrum measurements.  We recall that the index of
refraction of an ionized, magnetized plasma is birefringent. The left-
and right-circular polarizations of an electromagnetic wave passing
through such a plasma undergo different phase shifts, known as Faraday
rotation, such that the phase difference $\Delta\varphi$ of the light
becomes
\begin{equation}
  \Delta\varphi = \frac{e^2}{(m_ec^2)^2}\lambda^2\int 
n_e(s)B_{\parallel}(s)\D{s} \equiv \lambda^2\Phi,
  \label{eq:faraday}
\end{equation}
where $n_e$ is the electron density of the plasma, $B_{\parallel}$ is
the component of the magnetic field along the line of sight, $e$ and
$m_e$ are the electron charge and mass, $\lambda^2$ is the wavelength
of the incident light, and the integral extends along the line of
sight. Equation \ref{eq:faraday} defines the rotation measure $\Phi$.
Faraday rotation affects the linear components of the Stokes
parameters such that a polarized source with intrinsic Stokes $Q$ and
$U$, when viewed through a magnetized plasma, will have measured
Stokes parameters
\begin{equation}
  (Q + iU)_{meas} = e^{-2i\lambda^2\Phi}(Q + iU)_{intr}
  \label{eq:QUrot}
\end{equation}

The frequency structure induced by Faraday rotation differs from
normal smooth-spectrum foregrounds, and exhibits high covariance with
the high line-of-sight $k_{\parallel}$ modes, which are typically free
of synchrotron foreground emission \citep[e.g.][]{Morales2006,
  Morales2012, DelaySpectrum}. In fact, there is a nearly one-to-one
correspondence between a $\Phi$ mode and the $k_{\parallel}$ mode it
most infects, given by
\begin{equation}
  k_{\parallel} = 4\frac{H(z)}{c(1+z)}\Phi\lambda^2,
  \label{eq:k_infect}
\end{equation}
where $z$ is the redshift of the observation, $H(z)$ is the Hubble
parameter at that redshift, $\lambda$ is the wavelength of the
observation, and $\Phi$ is the rotation measure in
question.\footnote{A similar equation to this has been presented in
  two other papers, M13 and \citet{Pen2009}. Both of these contain
  algebraic mistakes, which are corrected in the formulation we present
  here. We thank Gianni Bernardi for pointing out these mistakes.}
At 164 MHz, the central frequency of one of the bands we present, a
typical rotation measure of $20\ {\rm {rad ~m}^{-2}}$ will infect
$k_{\parallel}$ values of around $0.25\ h\,{\rm Mpc^{-1}}$, well
outside the horizon limit for smooth-spectrum foregrounds on short
baselines ($k \approx 0.05\ h\,\mathrm{Mpc}^{-1}$ for $z=7.7$ and a
\mbox{30 m} baseline).

The power spectrum for 21cm EoR measurements is unpolarized, so the
frequency structure induced by Faraday rotation must leak into $I$
measurements through instrumental effects. We note that any instrumental
effect that leaks Stokes $Q$ or $U$ into $I$ is subject to the kind
of contamination we discuss here.  For PAPER, the particular form of
the dominant leakage comes about as follows.  Since PAPER has little
imaging capability in its maximum-redundancy configuration, we cannot
form Stokes parameters in the image plane, but rather, we combine
visibilities as if they were images, by defining
\begin{equation}
  \begin{pmatrix}V_I\\V_Q\\V_U\\V_V\end{pmatrix}
  = \frac{1}{2}
  \begin{pmatrix}
    1 &  0 & 0 &  1 \\
    1 &  0 & 0 & -1 \\
    0 &  1 & 1 &  0 \\
    0 & -i & i &  0
  \end{pmatrix}
  \begin{pmatrix}V_{xx}\\V_{xy}\\V_{yx}\\V_{yy}\end{pmatrix}
  \label{eq:Stokes},
\end{equation}
where $V_{xx}$, etc., are the linearly polarized, measured visibilities
for each frequency and time, and $V_I$, etc., are the ``Stokes
visibilities''.  The detailed expression for the leakage of Stokes $Q$,
$U$, and $V$ into $I$ due to wide-field beams has been explored elsewhere
\citep[e.g.,][]{shaw15}, and we defer a detailed study of this effect
for PAPER to future work.  Considerable simplification occurs in the
limit that the two feeds produce orthogonal responses to the two
electric field polarizations everywhere on the sphere, which is not
strictly achievable in practice \citep[see, e.g.,][]{carozzi09}, but is a
good approximation over most of the 45\arcdeg\ FWHM PAPER
field of view.  In this limit, expanding the first row of the matrix
and expressing the visibilities in terms of the beams $A_p$ for
polarization $p$, the baseline $\Vector{b}$, the unit vector
$\hat{s}$, and the intrinsic polarized signals $I$ and $Q$, we find
\begin{align}
\label{eq:BeamLeakage}
  V_I = \int (A_{xx} + A_{yy})Ie^{-2\pi i \Vector{b}\cdot\hat{s}\nu/c}\ \D{\Omega}
  \nonumber \\
   + \int (A_{xx} - A_{yy})Qe^{-2\pi i \Vector{b}\cdot\hat{s}\nu/c}\ \D{\Omega}.
\end{align}
The expression for $V_Q$ is similar, with $I$ and $Q$ interchanged.
Thus, if the primary beam of each element is not symmetric under rotations of
90$^\circ$, then the $I$ visibility will have a contribution from both
$I$ and $Q$. This provides the mechanism for the spectrally non-smooth
Faraday-rotated polarized emission to infect the typically unpolarized
21cm EoR power spectrum for PAPER.

\section{Data Processing}
\label{sec:DataProcessing}

The data used to create these results are nearly identical to those
presented in this paper's sister papers, P14 and J15.  We will review the processing steps presented
there, highlighting slight modifications.

PAPER's 32 antennas were arranged into a 4 row $\times$ 8 column grid during the
2011-2012 season (PSA32), when these data were taken. The east-west row spacing
was 30 m, and the north-south column spacing was 4 m. This choice of antenna
configuration maximizes baseline redundancy, achieving heightened
sensitivity \citep{PAPERSensitivity}, and allows for the redundant
calibration of visibilities \citep[e.g.][]{Liu2010}.

\begin{figure}
  %\plotone
  \includegraphics[scale=0.4]{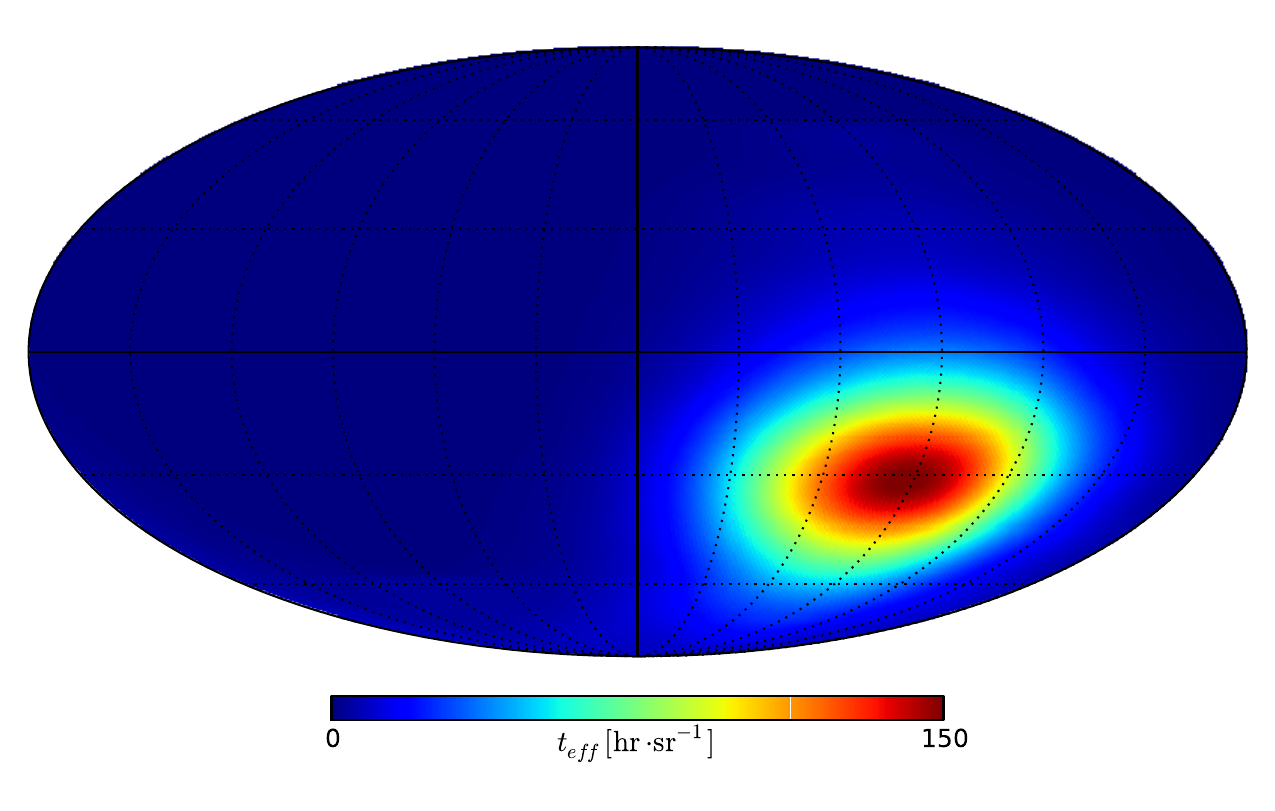}
  \caption{\label{fig:FOV} Effective integration time per pointing, as
    a function of position on the sphere. This metric is defined to
    give the total integration time when integrated over position on
    the sphere. The total field of view surveyed is 2.39 sr.  RA zero
    is at the center of the Mollweide projection, and increases to the
    right, wrapping around the sphere.  The maximum occurs in RA at
    approximately 6h, and in Dec at approximately -30\arcdeg, above
    the latitude of the array.}
\end{figure}

This data set consists of data taken continuously from Julian date
2455903 until 2455985, for a total of 82 days of observation.  The
effective integration time for any point on the sky is shown in Figure
\ref{fig:FOV}. Data are not considered when the sun is above -5$^\circ$
in altitude.

\begin{table*}\begin{center}
  \begin{tabular}{| c | c | c | c | c | c | c | c | }
    \hline
    & Center frequency & Total bandwidth & Channel width & Redshift & Collecting area & System temperature & Beam Leakage Ratio\\
    & $\nu_c$ [MHz] \ & $\Delta\nu$ [MHz] & $\delta \nu$ [kHz] & $z$ & $A_{\rm eff}$ [m$^2$] & $T_{\rm sys}$ 
[K] & $\mathcal{A}_-/\mathcal{A}_+$ (Eq. \ref{eq:def_beamleak}) \\
    \hline\hline
    Band \RNum{1} & 126 & 7.9 & 493 & 10.3 & 4.47 & 836 & $3.3\times10^{-3}$ \\
    \hline
    Band \RNum{2} & 164 & 9.4 & 493 & 7.66 & 5.80 & 505 & $2.2\times10^{-3}$ \\
    \hline
  \end{tabular}
  \caption{\label{tab:obs_params} Observational Parameters}
\end{center}\end{table*}

\subsection{Initial Processing}
\label{subsec:InitialProcessing}

We begin with an excision of radio-frequency interference (RFI), a
three-step process. First, we flag known frequency channels containing
nearly constant RFI, for example, the 137 MHz frequency bin that
contains the continuous signal from a constellation of communications
satellites.  Away from these regions, the fraction of data flagged is
$\ll 1\%$ of the total.  Next, we subtract adjacent time and frequency
channels from each other to cancel the bulk of the signal, and flag
$6\sigma$ outliers in the differenced data. Finally, we remove a
foreground model and flag $4\sigma$ outliers in the residual spectra.

Once RFI excision has taken place, we low-pass filter and decimate the
data in time and frequency, in the manner described in Appendix A of
P14. This reduces the data volume by roughly a factor of forty ---
from 1024 to 203 channels, and from integration times of 10 seconds to
34 seconds --- and preserves all celestial signal, including the EoR.

Next, we derive a fiducial calibration solution for a single day's
worth of data. We begin by solving for antenna-based gains and
electrical delays which enforce redundant measurements across
redundant baselines in the array. This reduces the calibration
solutions to a single, overall gain and delay per polarization of the
array, which we solve by fitting visibilities to a model of Pictor A,
correcting for the primary beam gain towards Pictor A, as in
\citet{Jacobs2013b}.  For sources at zenith, calibration errors are a
few percent. We apply this fiducial calibration solution to all data.

We develop a model of the smooth-spectrum foregrounds for each
integration and each baseline by constructing a spectrum of delay
components over the full available bandwidth (100-200 MHz) using a 1D
CLEAN algorithm. We constraint these CLEAN components to lie within
the entire horizon-to-horizon range with an additional extent of 15 ns
beyond either horizon.  This procedure does not affect high-delay
signal due to the EoR, but both removes foreground signal and
deconvolves by an uneven RFI flagging kernel. We subtract this model
from the data at each integration and baseline. This procedure is
described in P14 and J15.

Finally, we remove cross-talk, defined as an offset in visibilities
constant with time. For each baseline, and for each day, we subtract
the nightly average of the data, enforcing mean zero visibilities.

\subsection{Polarization Calibration}

To begin a discussion of polarization calibration, we summarize the
redundant calibration procedure we take. First, we take the set of
$xx$ and $yy$ visibilities and treat them like independent
arrays. Within each of these arrays, we solve for the $N_{\rm ant}$
antenna-based gains and $N_{\rm ant}$ antenna-based electrical delays
which force all baselines of a certain type to be redundant with a
fiducial baseline in each type. This leaves three calibration terms
per polarization to be solved for: an overall flux scale, and two
delays that set the three baseline types to the same phase
reference. We solve for these three by fitting the redundant
visibilities to a model of Pictor A \citep{Jacobs2013b}.

So far, we have made the reasonable assumption that $I$ dominates the
$xx$ and $yy$ visibilities. If we also assume that the gains and delays are
truly antenna-based, then we can apply the calibration solutions from
those visibilities to the $xy$ and $yx$ visibilities. This procedure
omits one more calibration term, which sets the $x$ and $y$ solutions
to the same reference phase: the delay between the $x$ and $y$
solutions, $\tau_{xy}$.

To solve for this delay, we minimize the quantity
\begin{equation}
  \chi^2 = \sum_{i,j}\left|V_{ij}^{xy} - V_{ij}^{yx}\exp\{-2\pi i 
\tau_{xy}\nu\}\right|^2,
\end{equation}
where the sum runs over antenna pairs $i,j$, finding the electrical
delay that minimizes $V_V$ in the least-squares sense. This
potentially nulls some signal in $V_V$, but we do not expect any
significant signal in $V$ at these frequencies. This method of
polarization calibration is similar to that presented in
\citet{Cotton2012}, but rather than maximizing $V_U$, we are
minimizing $V_V$.  By assuming that gains are antenna-based and that
the flux in $V_{xx}$ and $V_{yy}$ is dominated by unpolarized
emission, we need not correct for gain differences between $V_{xy}$
and $V_{yx}$ (the relevant gains having already come from the $xx$ and
$yy$ solutions).

\subsection{Averaging Multiple Days}
\label{sec:LSTBinning}

As a final excision of spurious signals (most likely due to RFI), on
each day, we flag outlying measurements in each bin of local sidereal time (LST). We use the
measurement of $T_{\rm sys}$ outlined in Section \ref{sec:Tsys} to
estimate the variance of each bin, and flag $3\sigma$ outliers.

If the data followed a normal distribution, consistent with pure,
thermal noise, then this procedure would flag around one measurement
in each frequency/LST bin, causing a slight miscalculation of
statistics post-flagging. To counteract this effect, we calculate the
ratio of the variance of a normal distribution, truncated at
$\pm3\sigma$ (97.3\%). We increase all errors in the power spectrum by
a factor of $1.03\approx 1/97.3\%$ to account for this effect.

We compute the mean of the RFI-removed data for each bin of LST and
frequency, creating a data set that consists of the average over all
observations for each LST bin, literally an average day. We continue
analysis on these data.

\subsection{Final Processing}

After visibilities are averaged in LST, a final round of cross-talk
removal is performed. Again, we simply subtract the time average
across LST from the data.

In the penultimate processing step, we pass the data through a
low-pass filter in time.  \citet{Parsons2009} and Appendix A of P14
describe the celestial limits of the fringe rate $f$ for drift-scan
arrays as $b_E\omega_\oplus\cos\lambda \le f \le b_E\omega_\oplus$,
where $b_E$ is the east-west component of the baseline,
$\omega_\oplus$ is the angular velocity of the Earth's rotation, and
$\lambda$ is the latitude of the observation. We filter the data in
time using a boxcar filter in fringe-rate space, defined as one on $0
\le f \le b_E\omega_\oplus$ and zero elsewhere. While this does null
some celestial emission (roughly the area between the south celestial
pole and the southern horizon), its effect is small, since the primary
beam heavily attenuates these areas of the sky. We null these fringe
rates as an additional step of cross-talk removal.

Finally, we combine the linearly polarized visibilities into Stokes
visibilities, as in Equation \ref{eq:Stokes}.

\subsection{System Temperature}
\label{sec:Tsys}

Once initial preprocessing has been completed, we take advantage of
nightly redundancy as a final check on the data. Since PAPER is a
transit array, measurements taken at the same LST on different nights
should be totally redundant. This redundancy allows us to measure the
system temperature via fluctuations in signal in the same LST bin from
day to day.

First, we compute the variance in each frequency and LST bin over all
nights of data $\sigma_{Jy}^2(\nu,t)$, and convert this variance into
a measure of the system temperature $T_{\rm sys}$. This measurement is
totally independent of the following power spectral analysis, and can
be used to quantify the level of systematic and statistical
uncertainty in the power spectra. It complements measurements of
$T_{\rm sys}$ from the uncertainties in power spectra in P14 and J15. The
variance computed in each frequency/LST bin $\sigma_{Jy}^2(\nu,t)$ is
converted into a system temperature in the usual fashion:
\begin{equation}
  T_{\rm sys}(\nu,t) = \frac{A_{\rm eff}}{k_B}\frac{\sigma_{Jy}}{\sqrt{2\delta\nu 
t_{\rm int}}},
  \label{eq:def_tsys}
\end{equation}
where $A_{\rm eff}$ is the effective area of the antenna, $k_B$ is the
Boltzmann constant, $\delta\nu$ is the channel width, and $t_{\rm int}$ is
the effective integration time of the LST bin.

\begin{figure}
  %\plotone
  \includegraphics[scale=0.45]{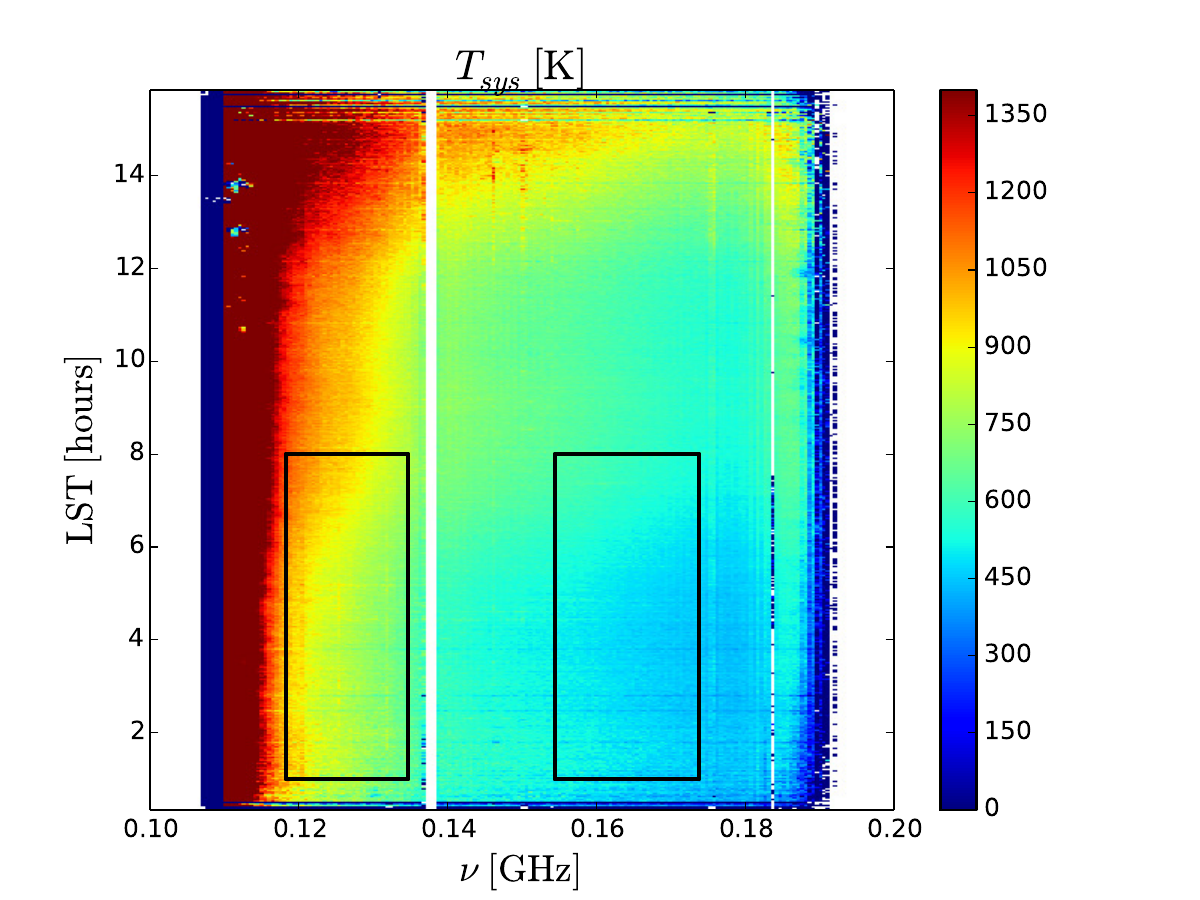}
  \caption{\label{fig:Tsys} System temperature in Kelvin as a function
    of local sidereal time (LST) and frequency $\nu$, calculated by
    Equation \ref{eq:def_tsys}.  Solid black boxes enclose the area
    used to compute the power spectrum.  The persistent RFI mentioned
    in Section \ref{subsec:InitialProcessing}, as well as the band
    edges, appear white in this figure.}
\end{figure}

Figure \ref{fig:Tsys} shows the measured system temperature for each
frequency and LST bin collected during the PSA32 observing season.  To
further summarize our data's variance, we average $T_{\rm sys}(\nu,t)$
over the time axis and frequency axis. The frequency-averaged system
temperature at center frequency $\nu_c$ may be computed as
\begin{equation}
  \langle T_{sys} \rangle_{\nu_c} (t)
  \equiv 
  \frac{\int_{\Delta\nu} W(\nu; \nu_c) T_{sys}(\nu, t)\ 
\D{\nu}}{\int_{\Delta\nu} W(\nu; \nu_c)\ \D{\nu}},
  \label{eq:Tsys_avg}
\end{equation}
where $W(\nu; \nu_c)$ is the window function in frequency used and the
integral spans the bandwidth $\Delta\nu$. For our analysis, we use a
Blackman-Harris window function \citep{BlackmanHarris}, chosen to
maximally suppress sidelobes. A similar expression may be written for
the time axis, where our window function is simply the number of
redundant samples in each frequency channel.

\begin{figure}
  %\plotone
  \includegraphics[scale=0.45]{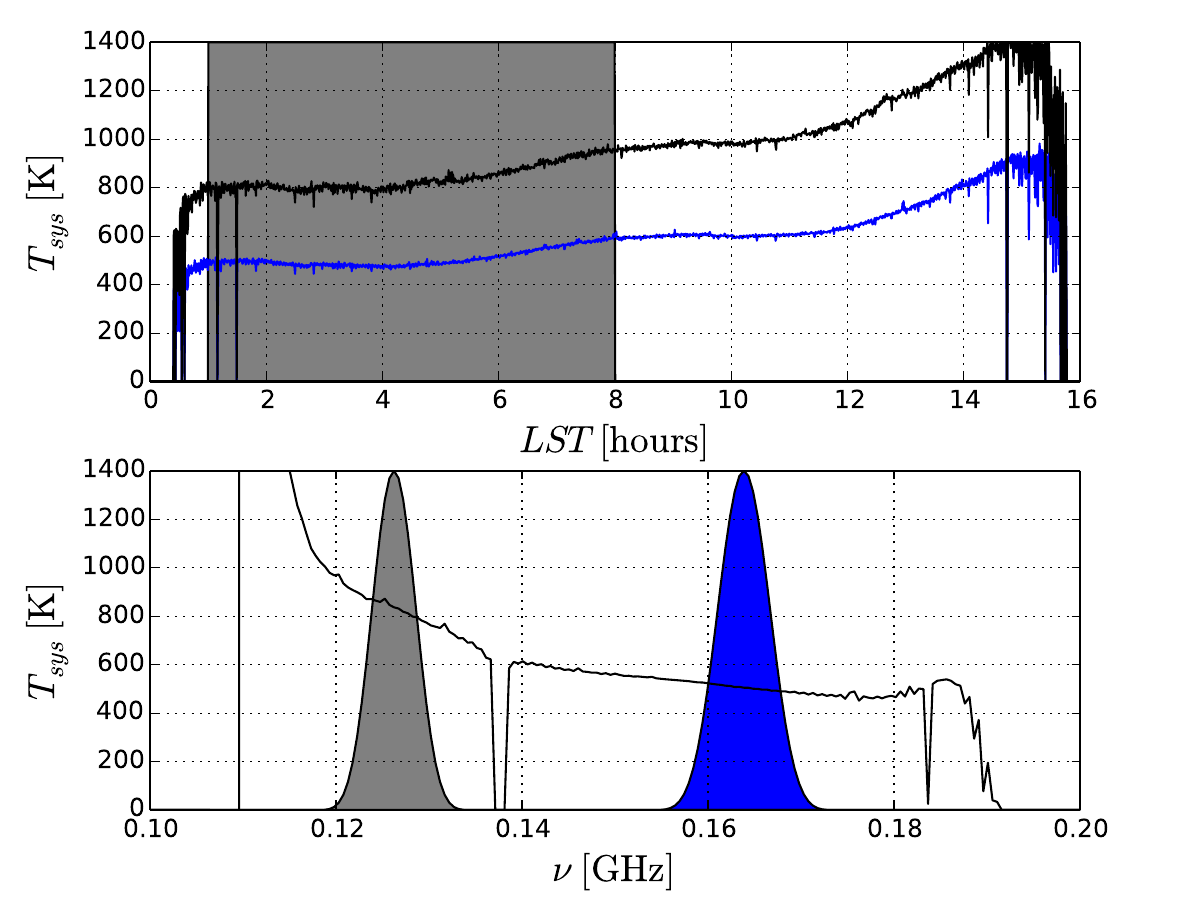}
  \caption{\label{fig:Tsys_lst} (Top Panel) Band-averaged system
    temperature (Equation \ref{eq:Tsys_avg}) as a function of local
    sidereal time for band \RNum{1} and \RNum{2} in black and blue,
    respectively. The shaded gray region indicates the range in local
    sidereal time used to compute the power spectra. (Bottom panel)
    Time-averaged system temperature, averaged over local sidereal
    times 1h00m to 8h00m. The shaded gray and blue regions show the
    window functions for Bands \RNum{1} and \RNum{2},
    respectively. The persistent RFI mentioned in Section
    \ref{subsec:InitialProcessing}, as well as the band edges, appear
    as zeros in this estimate of of the system temperature.}
\end{figure}

Figure \ref{fig:Tsys_lst} shows the system temperature averaged over
frequency and LST ranges used to compute the power spectra. Frequency-
and time-averaged system temperatures for the bands centered at 124 MHz
(hereafter Band \RNum{1}) and 164 MHz (Band \RNum{2}) are reported in
Table \ref{tab:obs_params}.

\subsection{Power Spectra}
\label{subsec:Results}

We compute the power spectrum using the delay spectrum approach
\citep{DelaySpectrum}. For short baselines, the delay transform of a
visibility, defined as
\begin{equation}
  \widetilde{V}(\tau) = \int V(\nu)e^{2\pi i \tau \nu}\ \D{\nu}
  \label{eq:DelayTransform}
\end{equation}
for visibility $V(\nu)$ and delay mode $\tau$, becomes an estimator
for $T(\Vector{k})$, the Fourier-transformed brightness temperature field. The
power spectrum may be computed from the delay-transformed visibilities
via the equation
\begin{equation}
  P(\Vector{k}) = \left(\frac{\lambda^2}{2k_B}\right)^2
  \frac{X^2Y}{\Omega\Delta\nu}\left|\widetilde{V}(\tau)\right|^2.
  \label{eq:Dspec}
\end{equation}
Here, $P(\Vector{k})$ is the three-dimensional power spectrum of 21 cm
emission, $\lambda^2/2k_B$ is the conversion from Jy to K, $\Omega$ is
the solid angle subtended by the primary beam, $\Delta\nu$ is the
bandwidth of the observation, $X^2Y$ is the factor converting
cosmological volume in $h^{-3} {\rm Mpc}^3$ to observed volume
$\Omega\Delta\nu$ (taken from Equations 3 and 4 in \citet{FOB}), and
$\widetilde{V}(\tau)$ is the delay-transformed visibility. The
$\Vector{k}$ modes are determined by the baseline vector and the
$\tau$ mode.

A subsequent covariance removal, described in detail in Appendix C of
P14, projects the delay-transformed visibilities into a basis in which
the covariance between two redundant baselines is diagonal, and then
computes the power spectrum from the projected delay spectra. This
procedure produces an estimate of the power spectrum for each LST bin
and baseline type. To measure the uncertainties in the time-dependent
power spectra, we bootstrap over both redundant baselines and LST
samples.

\subsection{Accounting for Ionospheric Effects}
\label{subsec:Ionosphere}

In P14 and J15, we ignored the effects of the ionosphere, since for
Stokes $I$ these are largely changes in source position induced by
refraction \citep[for a recent study, see][]{loi15} and these are
negligible on the large angular scales considered
\citep{vedantham15a,vedantham15b}.  However, both spatial and temporal
fluctuations in the Faraday depth of the Earth's ionosphere will have
a strong effect on polarized signal.  As the total electron content
(TEC) varies, it modulates the incoming polarized signal by some
Faraday depth that is a function of both the TEC for that time and
the strength of the Earth's magnetic field. Though we assume that
visibilities are redundant in LST for the purposes of averaging to
form the power spectrum, they do in fact have variations due to the
variable TEC of the ionosphere. Thus, averaging in LST could result in
attenuation of signal.  We are not able to directly image each day and
calculate the effects of ionosphere variations based on the properties
of celestial sources, but we are able to estimate the size of the
effect on the visibilities used in the analysis of the power spectrum.

\subsubsection{Characterizing Ionospheric Behavior over the Observing Season}
\label{subsubsec:characterize_Ionosphere}

Ionosphere data are provided by a number of sources.  We have used data
from the Center for Orbit Determination in Europe (CODE), whose global
ionosphere maps are available in IONosphere map EXchange format
(IONEX) via anonymous ftp. IONEX files from CODE are derived from
$\sim$200 GPS sites of the International Global Navigation Satellite
System Service (IGS) and other institutions. The time resolution of
CODE data is 2 hours, and the vertical TEC values are gridded into
pixels 5$^{\circ}$ across in longitude, and 2.5$^{\circ}$ in latitude.

We have used the core of the code provided from the
\ionFR\ package\footnote{ http://sourceforge.net/projects/ionfarrot/}
described in \citet{sotomayor-beltran13,sotomayor-beltran15} to access
the IONEX files.  As written, \ionFR\ provides the RM toward a given
RA as a function of UT at a given latitude (with the CODE two-hour
time resolution interpolated into hourly values). We have generalized \ionFR\ to calculate ionospheric RM values over a {\sc healpix} sphere with nside=16 (the maximum spatial resolution obtainable from an IONEX file) using a custom-built {\sc python} package {\tt radionopy}\footnote{\url{https://github.com/jaguirre/radionopy}}.

For each day of the PSA32 campaign, we obtained the relevant IONEX
file.  For a fixed LST, we found the UT corresponding to that LST at
transit for the PSA32 site for each day of observation.  Using the
data in an IONEX file and {\tt radionopy}, we were able to calculate the
vertical TEC and geomagnetic field (and hence RM) values over the
entire hemisphere observed by PAPER. The result is a map of 
$\Phi(\Omega)$ giving the ionospheric RM $\Phi$ induced for sources in
any direction; an example is shown in Figure~\ref{fig:RM_snap}.

\begin{figure}
\animategraphics[controls,loop,scale=0.45]{3}{RM_2012-02-13_UT}{0}{23}
\caption{Ionospheric RMs over the PSA32 site during 2012 February 13. Times are shown in UT (South African Standard Time is UT+2). The plot is shown in local horizon coordinates, with zenith at the center and the horizon around the edges.  Note the large, smooth variation over the FoV. All of these RMs affect the measured visibilities, making accurate polarization calibration extremely difficult. The reader can view an animated hour-by-hour sequence by viewing this figure in {\sc adobe acrobat reader}.}
\label{fig:RM_snap}
\end{figure}
 
\begin{figure}
\includegraphics[scale=0.45]{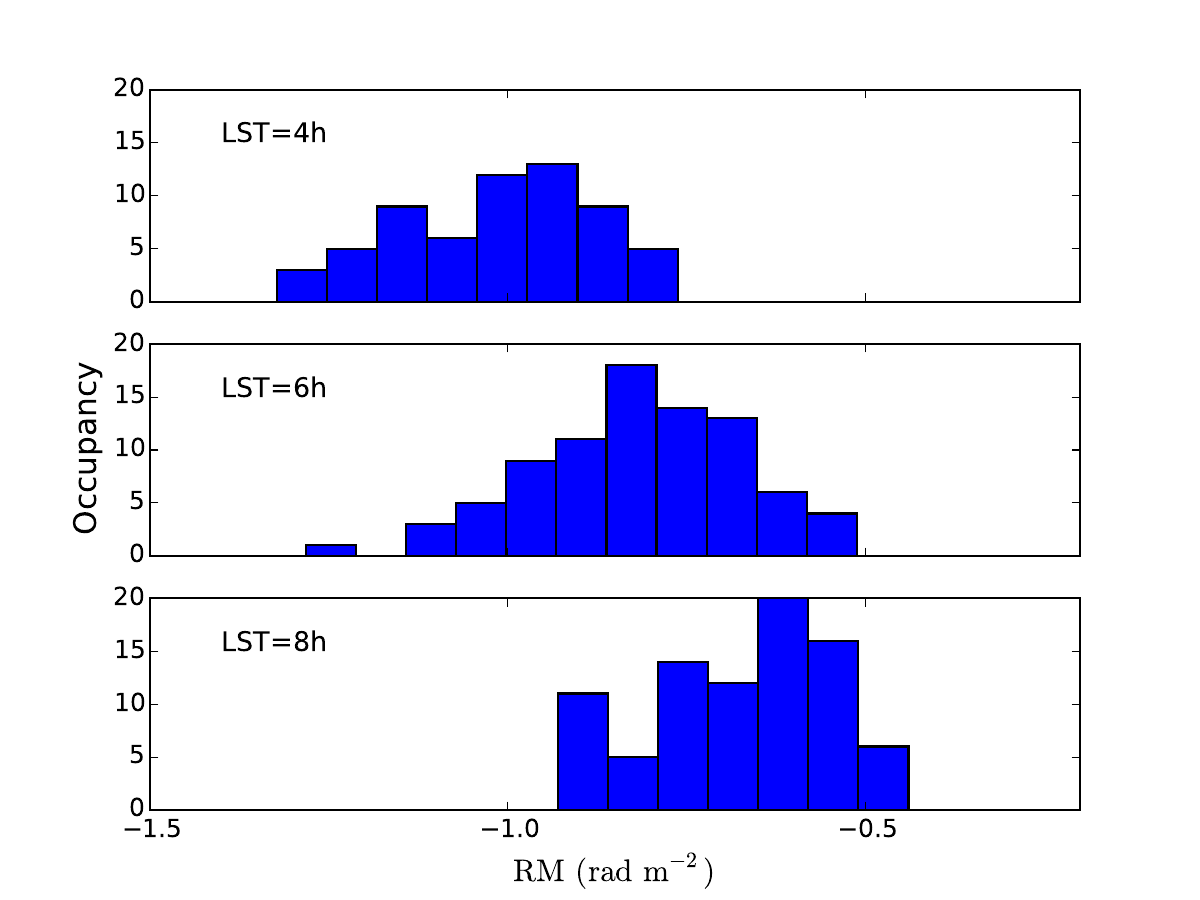}
\caption{From top to bottom, a histogram of the zenith ionospheric RMs
  over the season, for the transit of LSTs 4h, 6h and 8h. The large
  variance at any given LST corresponds to a large attenuation of
  polarized signal, as discussed in
  Section~\ref{subsubsec:calculate_Ionosphere}.}
\label{fig:RMspread}
\end{figure}

To give an indication of the time variability of the RM, as coadded
into the power spectrum for that LST (recall Section
\ref{sec:LSTBinning}), we calculated for each night of observation the
zenith RM when three LSTs (4h, 6h, and 8h) were at transit. These are
shown in Figure~\ref{fig:RMspread}.  Over the season, there is a large
spread of ionospheric RMs for each LST. As the relevant phase shift of
the Faraday-rotated spectrum is $\Phi \lambda^2$, this clearly varies
by more than a radian over the range of days coadded.  This will lead
to a large attenuation of polarized power during LST-binning; we
calculate this attenuation in the next section. Also note that there
is a decrease in the average magnitude of the RM as LST increases. This is
expected, given the strong correlation between the day / night cycle
and TEC values \citep{RadicellaAndZhang, Tariku}, and given that for
this observing season, LST=4h corresponds to observing times shortly
after sunset, while LST=8h is always well into the night.

\subsubsection{Calculating Ionospheric Attenuation of Polarized Signal}
\label{subsubsec:calculate_Ionosphere}

The effect of the ionosphere requires a modification of equation
\ref{eq:BeamLeakage} to account for the effect of Equation
\ref{eq:QUrot}, which following the formalism of \citet{HBS1}, is of
the form
\begin{align}
\label{eq:V_I_Ionosphere}
  V'_I = \int \D{\Omega} \, e^{-2\pi i \Vector{b}\cdot\hat{s}\nu/c} [(A_{xx} + A_{yy}) I + \nonumber \\
   (A_{xx} - A_{yy})(Q \cos\phi + U \sin\phi)] %\nonumber \\
   %\equiv {\cal v}_I + {\cal v}_P
\end{align}
where $\phi=2 \Phi(\Omega) \lambda^2$ and $\Phi(\Omega)$ represents
the spatial distribution of ionospheric rotation measures at the time
of observation.  As we have seen, $\Phi(\Omega)$ is a slowly varying
function over the PAPER beam.

If we assumethat  $\Phi(\Omega)$ is spatially constant, Equation
\ref{eq:V_I_Ionosphere} can be rewritten
\begin{eqnarray}
    V'_I & = & \int \D{\Omega} \, e^{-2\pi i \Vector{b}\cdot\hat{s}\nu/c} (A_{xx} + A_{yy}) I \nonumber \\
   & + & \cos\phi \int \D{\Omega} \, e^{-2\pi i \Vector{b}\cdot\hat{s}\nu/c} (A_{xx} - A_{yy}) Q  \nonumber \\ & + & \sin\phi \int \D{\Omega} \, e^{-2\pi i \Vector{b}\cdot\hat{s}\nu/c} (A_{xx} - A_{yy}) U \nonumber \\
   & \equiv & {\cal V}_I + \cos\phi {\cal V}_Q + \sin\phi {\cal V}_U
\end{eqnarray}
and we can write the LST-averaged visibility as the Faraday
rotation-weighted sum of otherwise redundant visibilities:

\begin{equation}
  \widehat{V} = \frac{1}{N}\sum_i {\cal V}_I + \cos\phi_i {\cal V}_Q + \sin\phi_i {\cal V}_U,
  \label{eq:ionosphere_avg}
\end{equation}
where $\phi_i$ is the zenith ionospheric Faraday depth from day $i$
and the other terms are the redundant component of the visibilities.
Note that ${\cal V}_I$ does not provide contributions at high $k$ to
the power spectrum due to the assumption of smooth-spectrum
foregrounds, and we are concerned only with the leakage due to ${\cal
  V}_Q$ and ${\cal V}_U$.  These visibilities' contribution to the
power spectrum for given LST will average approximately as
\begin{equation}
  \widehat{P} \propto |\widehat{V}|^2
  = \frac{1}{N^2}\left(\sum_{i,j}e^{-2i(\Phi_i-\Phi_j)\lambda^2}\right)|V|^2 
\equiv 
  \varepsilon |V|^2.
\end{equation}
The sum may be rewritten in terms of the $i=j$ components and the
$j>i$ component:
\begin{equation}
  \sum_{i,j}e^{-2i(\Phi_i-\Phi_j)\lambda^2} = N +
  2\sum_{i>j}\cos\left\{2(\Phi_i-\Phi_j)\lambda^2\right\}.
  \label{eq:attenuate}
\end{equation}
In the limit where all values of $\Phi_i$ are equal, this second term
becomes $N(N-1)/2$, the number of $i,j$ pairs with $i>j$, showing that
with no daily fluctuations in ionospheric Faraday depth, there is no
change in the signal.

To estimate the level of ionospheric attenuation, we calculated
$\varepsilon$ for the 3 LST transits described in the section
above. For the observed distribution of RM at LST = 4h, 6h, 8h, the standard deviations of RM are $\sigma_\Phi = 0.30, 0.24, 0.20$ rad m$^{-2}$ and 
the average attenuation was calculated to be
$\varepsilon=0.42, 0.41, 0.48$ in Band II (164 MHz) and $\varepsilon=0.07, 0.06, 0.11$ in Band I (126 MHz), for each LST, respectively. 
To obtain some measure of the uncertainty in
this value, we generated 10,000 realizations of the attenuation factor
for an 82 day integration, drawing randomly from the RM distributions
shown in Figure~\ref{fig:RMspread}, and found the average simulated
value to be $0.43 \pm 0.06$ (Band II) and $0.07 \pm 0.05$ (Band I).  
(Note that while we have neglected it,
spatial variation in the TEC would tend to increase
the attenuation.)  The net effect is that polarized leakage into Stokes
$I$ is notably decreased by this averaging process, as are the
observed signals in Stokes $Q$ and $U$.  A more sophisticated model of this leakage and attenuation is the subject of future work.

\section{Results}
\label{sec:Results}

\subsection{Features of the Power Spectra}

\begin{figure*}
  %\plotone
  %\centering
  \hspace{-0.7in}
  \includegraphics[scale=0.7]{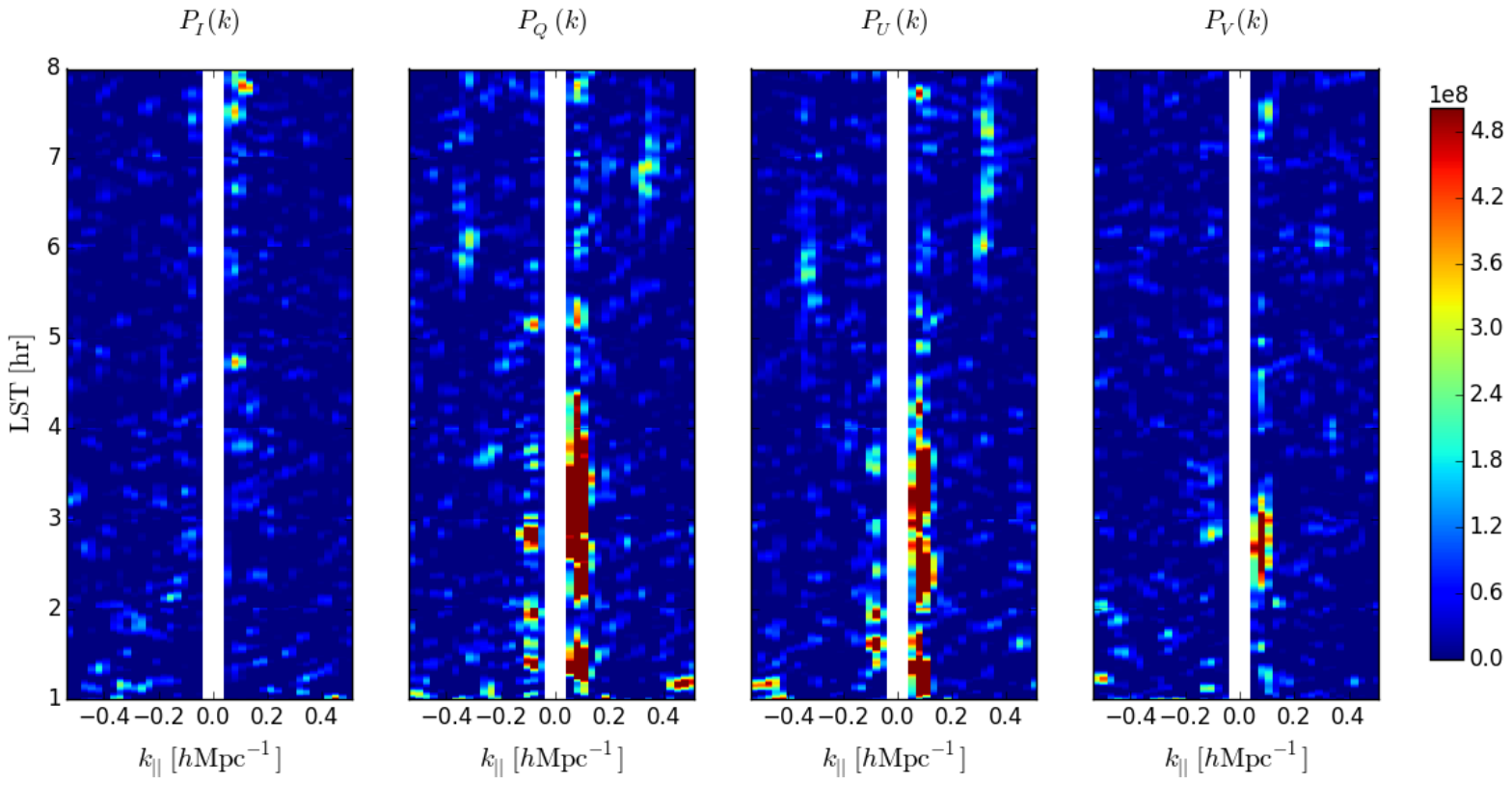}
  \caption{\label{fig:pq_vs_t} Power spectra $P(k)$ from Band \RNum{2}, in units of ${\rm
      mK}^2\,h^3\,{\rm Mpc}^{-3}$, shown as a function of $k_{\parallel}$ and
    local sidereal time (LST) in hours, which is the same for all panels. $k_{\parallel}$ modes within the horizon for
    the 30 m baselines used are masked. From left to right, the panels show $P_I$, $P_Q$, $P_U$ and $P_V$.
    Features in these power spectra are discussed further in the text.}
    \vspace{14pt}
\end{figure*}

\begin{figure}
  \includegraphics[scale=0.45]{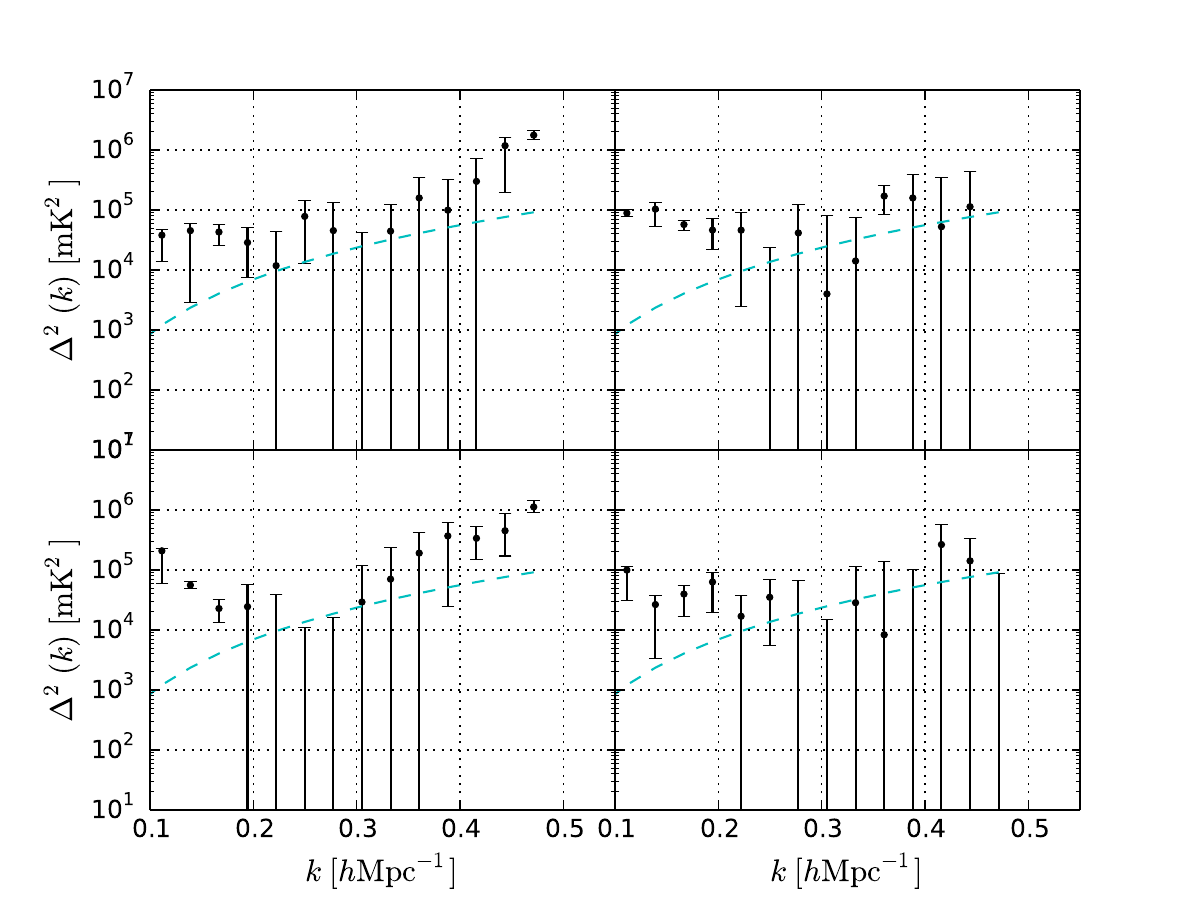}
\includegraphics[scale=0.45]{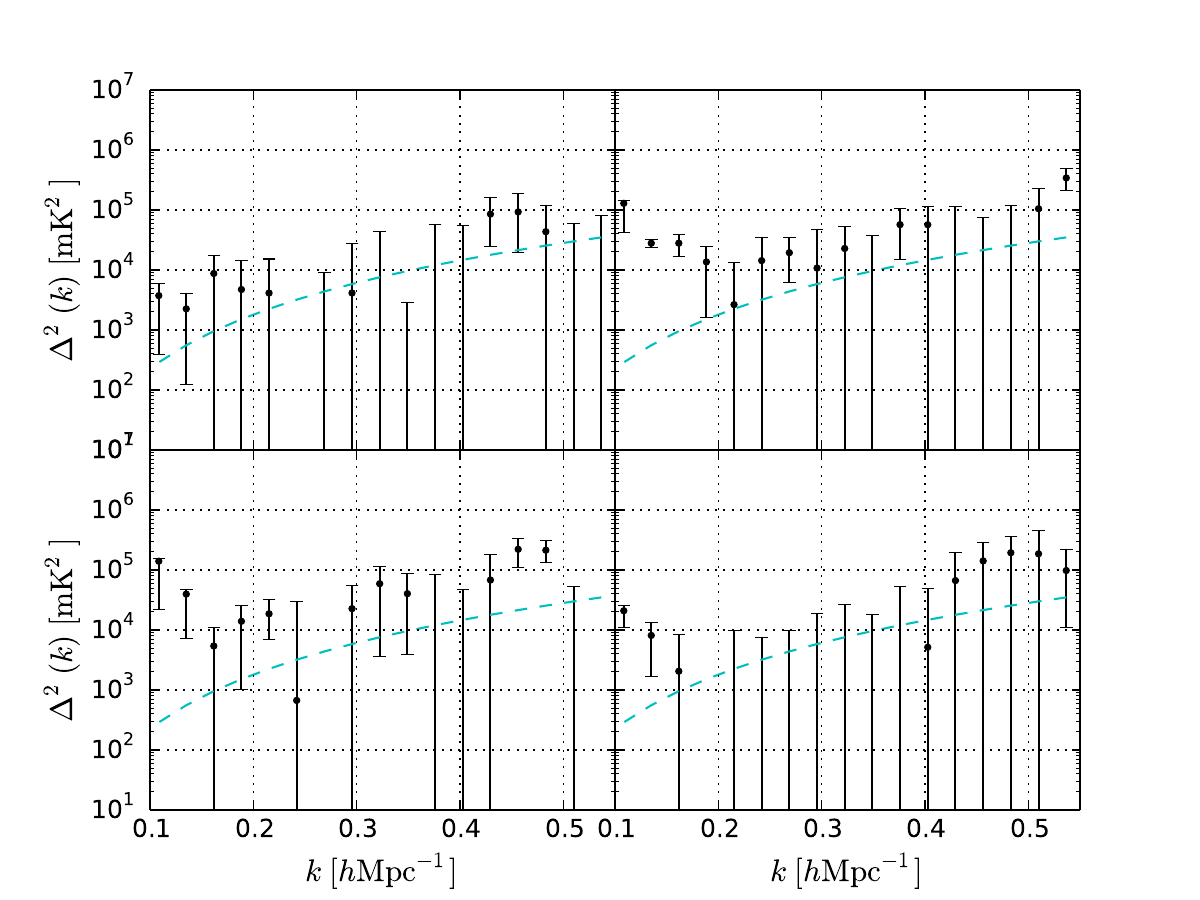}
  \caption{\label{fig:pspec} {\it Top:} Spherically-averaged power spectra for
    the four Stokes parameters for Band \RNum{1}. Stokes $I$ is in the top left panel, $Q$ in the
    top right, $U$ in the lower left, and $V$ in the lower right. Error bars show 98\% confidence
    intervals.  Dashed, cyan lines indicate the theoretical level of
    thermal fluctuations, with $T_{sys}$ calculated in Section
    \ref{sec:Tsys}.
  {\it Bottom}: Same as the above, for Band \RNum{2}.}
\end{figure}

Figure \ref{fig:pq_vs_t} shows the power spectra resulting from the above analysis in $I$, $P \equiv Q+iU$ and $V$ as a function of LST for Band \RNum{2}.  
Figure \ref{fig:pspec} shows the LST average of these power spectra for Stokes $I$, $Q$, $U$ and $V$ in Bands \RNum{1} and \RNum{2}. Uncertainties in these power spectra are the bootstrap errors described in Section \ref{subsec:Results}.   
The expected thermal noise sensitivity using the $T_{sys}$ computed from Section \ref{sec:Tsys} and the sensitivity calculations of \citet{PAPERSensitivity} and \citet{Pober2014} are shown as dashed, cyan lines.

In Figure \ref{fig:pq_vs_t}, we see that Band II Stokes $P_I$ and $P_V$ are largely devoid of 
features in LST. In the $P_Q$ and $P_U$ panels of Figure \ref{fig:pq_vs_t}, there are two major features worth noting.  The first is the excess of emission at $|k_{\parallel}| \lesssim 0.2\ h{\rm Mpc}^{-1}$, between right ascension 1h00m and
4h30m. That the excess exceeds the power in the corresponding $k$-bins of $P_I$ indicates that it is not due to leakage from Stokes $I$.  It roughly corresponds in RA with the diffuse,
polarized power shown in B13, which is evident between RA 22h and 3h, and takes its minimum value at around RA 5h.  The low RM reported in B13 would mean that this emission would appear in our power spectra at low $k$, as observed.  Whether this emission should be identified with the B13 features would require a true imaging analysis to establish.  

The second feature of Figure \ref{fig:pq_vs_t} that we will comment on is the tracks of excess power in Q and U between RA 6h and 8h, with $|k_\parallel| \approx 0.35 h{\rm Mpc}^{-1}$. Features such as these could be produced by polarized point sources of an apparent polarized flux of $\sim1.5$~Jy observed through a Faraday screen with $\Phi \sim 60$~rad m$^{-2}$.  Accounting for the ionospheric attenuation factor $\varepsilon$, the intrinsic polarized flux at 164 MHz would need to be $\sim4$~Jy, and larger still if the sources did not pass through zenith  (corresponding to RA 6h52m Dec -30\arcdeg).  While the features appear to transition between showing power in $P_Q$ and $P_U$, which would be expected for a intrinsically polarized point source undergoing parallactic rotation, the rotation between Q and U is faster in LST than expected for a celestial source.   Nevertheless, to see whether there is a plausible bright source in this region, we refer to the results of \citet{hurley-walker14}.  This survey covered 6100 square degrees, including the region above; excluding the brightest known sources (e.g., Pictor A, Fornax A), which do not match the position of the features, there are 22 sources with Stokes $I >10$~Jy in the primary beam of PAPER ($-52.5\arcdeg < \rm{Dec} < -7.5\arcdeg$) in that RA range, with the brightest being 28 Jy at 180 MHz.  While it is not impossible that a source could be $>10\%$ polarized, the typical polarization fraction at these frequencies is much lower \citep{lenc16}.  In addition to the point sources of  \citet{hurley-walker14}, we also note that the Galactic Plane passes through zenith in this RA range.  While the RM in the Plane is expected to be low, and the spectral smoothness of the primary beam is expected not to scatter power to high $k_\parallel$ \citep{kohn16}, it cannot be ruled out the strong Stokes $I$ power in the Plane couples to the instrument and produces these features.   Other unidentified instrument systematics, which need not be constant in time as a result of changes in the instrument over the observing season, may also be responsible.  However, using jackknife tests, for example, to test for this is difficult, because the feature is only 5.1$\sigma$, and thus unless the feature is highly localized in some portion of the data, partitioning the data would degrade the significance to $\sim3\sigma$.  We acknowledge that this anomaly merits further investigation, and that definitive identification as a celestial source requires a polarized imaging survey.

In Figure \ref{fig:pspec}, the Stokes $I$ power spectra reproduce those of P14 and J15, as expected.  
Stokes $V$ is very nearly consistent with zero, except for the slight {\it negative} excess in the power spectrum at $k\sim 0.15$ (apparent as the missing point in the semi-log plot), which is clearly due to the bright (in absolute magnitude) feature at $0.05\,<\,k\,<\,0.15$ in the $P_V$ panel of Figure \ref{fig:pq_vs_t}.  It is not possible for astrophysical $V$ emission to produce a negative {\it power spectrum}, and thus this must represent a systematic artifact.  It is evident that Band I is less consistent with zero than Band II (for all Stokes parameters), with the excess particularly notable for $k<0.2$.  These features were noted by J15 for Stokes $I$ in the lower frequency band, and attributed to poorer cleaning of foreground power at lower frequencies and near the edge of the ``wedge''.

\subsection{Polarized Leakage into the EoR Power Spectrum}

As discussed in Section \ref{sec:PolarizationReview}, the dominant
form of leakage of polarized power into $P_I$ in the analysis of the PAPER power
spectrum comes from $Q \to I$ due to the primary beam
ellipticity, as given in Equation \ref{eq:BeamLeakage}.  In general,
calculating the fractional power leakage
\begin{equation}
 \xi = \frac{|V_Q|^2}{|V_I|^2}
\end{equation}
depends on knowing both the primary beam and the sky.  We have done
simulations to determine this factor in M13, 
but here we also develop an
approximation that allows us to estimate $\xi$ knowing only the
primary beam, under some assumptions about the source distribution on
the sky.

We approximate $I$ and $Q$ as Gaussian, random fields with mean zero
(as an interferometer, PAPER is insensitive to the mean value in any
case). We also assume that $\langle Q^2 \rangle \approx p^2 \langle
I^2 \rangle$, that is, there is an average polarization fraction $p$
of Stokes $I$ relative to $Q$, where $p\ll1$ \citep[which is indeed true at
higher frequencies;][]{Tucci2012}.  Under these assumptions, we can
write the square of the visibility $V_I$, proportional to the measured
$I$ spectrum, as
\begin{equation}
  |V_I|^2 = \mathcal{A}_+\otimes P_I + \mathcal{A}_-\otimes P_Q,
  \label{eq:VIsquared}
\end{equation}
where $P_I$ and $P_Q$ are the true power spectra of $I$ and $Q$, and
$\otimes$ denotes a convolution. The weighting factors
$\mathcal{A}_\pm$ are the contributions in power from the summed and
differenced primary beams, defined as
\begin{equation}
  \mathcal{A}_\pm \equiv \int|A_{xx} \pm A_{yy}|^2\ \D{\Omega}.
  \label{eq:def_Apm}
\end{equation}
A similar expression to Equation \ref{eq:VIsquared} can be written for
$V_Q$, with an interchange of $I$ and $Q$.  Then 
\begin{equation}
  \xi = \frac{p^2\mathcal{A}_+ + 
\mathcal{A}_-}{\mathcal{A}_+ + p^2\mathcal{A}_-}
  \approx \frac{\mathcal{A}_-}{\mathcal{A}_+}.
  \label{eq:def_beamleak}
\end{equation}
This ratio, when multiplied by the measured $P_Q$ can be used as a
metric to characterize the level of polarization leakage present in a
measurement of $P_I$. Table \ref{tab:obs_params} shows the value of
this ratio for the PAPER beam \citep{Pober2012} for the two bands.
The ratio can also be calculated from simulations, and on average
agrees well with the simpler estimate of Equation \ref{eq:def_beamleak}.

The level of leakage predicted by Equation \ref{eq:def_beamleak} shows
that in the lowest $k_{\parallel}$ bins, the $I$ power spectrum of P14
and J15 cannot be dominated by $Q\to I$ leakage.  The levels of
polarized leakage are, to order of magnitude, $10^3\ {\rm mK}^2$ in
Band \RNum{1}, and $10^2\ {\rm mK}^2$ in Band \RNum{2}.  The lack of a
detection of polarized power (Stokes $Q$ and $U$) in Figure
\ref{fig:pspec} is consistent with what is known about polarized
emission, combined with the attenuation factor from the ionosphere in
Section \ref{subsubsec:calculate_Ionosphere}. Because we are considering only $k_\parallel>0.15~h~\rm{Mpc}^{-1}$, we assume the
contribution of low-RM diffuse emission is negligible (a result corroborated by \citet{kohn16}, who probe a much greater range of $k$-values over a comparatively short integration with the PAPER polarized imaging array).  For point
sources, a polarized fraction $p_{rm 150~MHz} = 0.006$ (lower than
that assumed in M13) would be consistent with the single reported
point source in B13 and an interpretation of the results of
\citet{asad15}.  When combined with the ionospheric attenuation, this
would produce a power spectrum below our limits, and leakage well
below the excess present in the lowest $k_{\parallel}$ bins of the
$P_I$ power spectra in P14 and J15. Indeed, if these levels are
typical, leakage would be below the levels reported in the more recent
PAPER-64 results of \citet{ali15}, albeit under different ionospheric
conditions.

\section{Conclusion}
\label{sec:Conclusion}

We have presented the first limits on the power spectra of all four
Stokes parameters in two frequency bands, centered at 126 MHz
($z=10.3$) and 164 MHz ($z=7.66$).  These data come from from a
three-month observing campaign of a 32-antenna deployment of PAPER.
These power spectra are processed in the same way as the results on the unpolarized
power spectrum have been reported at $z=7.7$
\citep{Parsons2014} and $7.5 < z < 10.5$ \citep{Jacobs2014}.  We do
not find a a definitive detection of polarized power.  The limits are
sufficiently low, however, that the level of polarized
leakage present in previous PAPER measurements must be less than 100
mK$^2$ at $k_{\parallel} \sim 0.2\ h{\rm Mpc}^{-1}$, below the excess
found in those measurements.

We additionally find that that the variation in ionospheric RM is
sufficient to attenuate the linearly polarized emission in these
measurements by factors between 2.5 and 16, depending on the observing frequency.  Combined with a lower
polarized fraction for point sources than assumed in
\citet{Moore2013}, this is consistent with our nondetection of
polarized power.

PAPER in the grid array configuration presented here is incapable of
creating the images with high dynamic range needed to isolate polarized
emission.  Future work will explore in more detail the effect of the
ionosphere and the full degree of depolarization present in long EoR
observing seasons, as well as the full effects of polarized leakage
from wide-field beams and polarization calibration.

\acknowledgements

PAPER is supported by grants from the National
Science Foundation (NSF; awards 0804508, 1129258,
and 1125558). ARP, JCP, and DCJ would like to acknowledge
NSF support (awards 1352519, 1302774, and
1401708, respectively). JEA would like to acknowledge a
generous grant from the Mount Cuba Astronomical Association
for computing resources. We graciously thank
SKA-SA for on-site infrastructure and observing support.

\bibliography{paper}

\end{document}